\documentclass[twocolumn,pra,longbibliography]{revtex4-1}
\usepackage{graphicx}
\usepackage{float}
\usepackage{dcolumn}
\usepackage{bm}
\usepackage{color}
\usepackage{amsmath}
\usepackage{amssymb}
\usepackage{amsfonts}
\usepackage{esint}
\usepackage{times}
\usepackage{xcolor}
\usepackage{phaistos}
\usepackage{braket}
\usepackage{comment}
\usepackage{multirow}

\usepackage{booktabs}

\usepackage{pgfplots}

\usepackage[colorlinks,linkcolor=blue,anchorcolor=blue,urlcolor=blue,urlcolor=blue,citecolor=blue]{hyperref}

\begin{document}

\title{Fast CZ gate in hybrid fluxonium-transmon systems with tunable couplers}

\author{Peng Xu$^{1,2,3,4}$}\email{pengxu@njupt.edu.cn}
\author{Yunlong Wang$^{1}$} 
\author{Yuechen Mu$^{1}$}  
\author{Ling Jiang$^{2}$} 
\author{Peng Zhao$^{5}$}\email{shangniguo@sina.com}
\author{Shengjun Wu$^2$}\email{sjwu@nju.edu.cn}
\author{Xiaohong Yan$^6$}\email{yanxh@njupt.edu.cn}
\affiliation{$^1$ Institute of Quantum Information and Technology, Nanjing University of Posts and Telecommunications, Nanjing, 210003, China \\
$^2$ National Laboratory of Solid State Microstructures and School of Physics, Collaborative Innovation Center of Advanced Microstructures, Nanjing University, Nanjing 210093, China.\\
$^3$ Jiangsu Key Laboratory of Quantum Information Science and Technology, Nanjing University, Suzhou, 215163 Jiangsu, China\\
$^4$ State Key Laboratory of Quantum Optics Technologies and Devices, Shanxi University, Taiyuan, 030006,China\\
$^5$ Quantum Science Center of Guangdong-Hong Kong-Macao Greater Bay Area, Shenzhen 518045, China\\
$^6$ School of Science, Nanjing University of Posts and Telecommunications, Nanjing 210023, China}

\begin{abstract} 

Hybrid superconducting architectures combining different types of qubits offer a promising platform for exploiting their complementary advantages, yet high-fidelity entangling gates remain challenging because of strong nonlinearities and residual qubit-qubit interactions. Here, we propose a high-fidelity controlled-Z (CZ) gate for a hybrid circuit comprising a fluxonium qubit, a fixed-frequency transmon qubit, and a flux-tunable transmon coupler. By modulating only the external magnetic flux applied to the coupler, the qubit-qubit interaction is dynamically engineered for conditional-phase accumulation while the residual interaction is suppressed at idle, mitigating spectator-induced errors. We employ a low-dimensional Fourier-cosine pulse parameterization and a physically motivated cost function to independently suppress conditional-phase errors and leakage from the computational subspace. Numerical simulations demonstrate that a microwave-free CZ gate can be realized within $25 \mathrm{ns}$, with an average gate fidelity exceeding $99.99\%$ and leakage below $10^{-5}$. Using experimentally relevant superconducting-qubit parameters and accounting for decoherence, the proposed scheme maintains a CZ-gate fidelity of approximately $99.9\%$. We further extend the analysis to larger coupled architectures and find that the CZ-gate infidelity remains below $10^{-4}$ in the presence of spectator qubits. These results establish single-parameter flux control as a simple and robust approach for realizing high-fidelity entangling gates in heterogeneous superconducting quantum architectures.

\end{abstract}

\maketitle

\section{Introduction}

Superconducting quantum circuits have become one of the leading platforms for quantum information processing owing to their excellent scalability, strong controllability, and compatibility with modern microfabrication technologies~\cite{Krantz2019,Blais2021}. Continuous improvements in materials, fabrication techniques, and circuit design have significantly prolonged qubit coherence times while enabling quantum gates with fidelities exceeding the fault-tolerance threshold~\cite{NSRhfyu2025}. Consequently, superconducting quantum processors containing tens to hundreds of qubits have been successfully demonstrated~\cite{FMeiPRL1251605032020, GoogleQuantumAINature6389202025, PhysRevLett1352606012025, HQianScience3909302025, arXiv2607121182026, arXiv2602212932026}, making superconducting circuits one of the most mature candidates for large-scale quantum computation.

Among various superconducting qubit modalities, the transmon qubit~\cite{Koch2007} has become the dominant building block for current quantum processors owing to its continually improving coherence times~\cite{SiddiqiNatRevMater68752021, MPBlandNature6473432025}, well‑established control technology, and excellent reproducibility~\cite{JVanDammeNature634742024}. More recently, the fluxonium qubit has also attracted considerable attention owing to its large anharmonicity and the possibility of achieving millisecond coherence times~\cite{Manucharyan2009, Nguyen2022, ASomoroffPhysRevLett1302670012023} and high-fidelity two-qubit gates~\cite{LDingPRX130310352023, HZhangPRXQuantum50203262024, WJLinPQXQuantum60103492025}. These complementary characteristics naturally motivate hybrid superconducting architectures that combine different qubit modalities within a single device~\cite{ACianiPhysRevRes40431272022, LHeunischArXiv250809267, NDDimitrovPhysRevAppl260140042026}, allowing the advantages of each qubit type to be simultaneously exploited.

Despite these advantages, implementing high-fidelity two-qubit entangling gates in hybrid superconducting systems remains a significant challenge. Compared with homogeneous transmon architectures, hybrid circuits involve multiple characteristic frequencies, different anharmonicities, and more complicated interaction spectra. As a consequence, unwanted residual couplings, state leakage, and parasitic conditional frequency shifts become increasingly difficult to suppress, thereby limiting both gate speed and achievable fidelity.

Flux-tunable couplers have emerged as an effective solution for dynamically engineering qubit-qubit interactions while suppressing unwanted static coupling~\cite{Chen2014, FeiYanPRApp100540622018, Mundada2019, HGotoPhysRevAppl180340382022, DLCampbellPRApp190640432023}. By tuning the coupler frequency through an externally applied magnetic flux, the effective interaction between neighboring qubits can be continuously switched from nearly zero in the idle configuration to a strong interaction during gate execution. Such tunable-coupler architectures have enabled a variety of high-fidelity entangling gates \cite{BFoxenPRL1251205042020, YuanXuPhysRevLett1252405032020, JStehlikPhysRevLett1270805052021, YoungkyuSungPhysRevX110210582021, RuiLiPhysRevX140410502024}.

Nevertheless, most existing tunable-coupler protocols have primarily focused on homogeneous transmon systems. Relatively little attention has been devoted to hybrid architectures involving fluxonium qubits, where the substantially different energy spectra and nonlinearities introduce additional challenges for gate design. Furthermore, many optimal-control approaches rely on high-dimensional pulse optimization, making the resulting control waveforms difficult to interpret physically and potentially challenging to calibrate experimentally.

In this work, we propose a high-fidelity controlled-Z (CZ) gate in a hybrid superconducting architecture consisting of a fluxonium qubit, a fixed-frequency transmon qubit, and a flux-tunable transmon coupler. By applying an external magnetic flux to the tunable coupler, its frequency is dynamically adjusted, which in turn reshapes the effective interaction landscape of the computational qubits. In particular, the coupler-mediated ZZ interaction can be selectively enhanced during CZ operations to generate the required conditional phase, while remaining strongly suppressed away from the interaction region, thereby reducing unwanted static ZZ coupling and spectator-induced crosstalk \cite{Mundada2019, RJSchoelkopfNature4602009, SPForsarXiv2408154022024}. The proposed single-control strategy thus enables both efficient entangling-gate implementation and suppression of parasitic interactions, providing a simple and robust approach to realizing fast and high-fidelity CZ gates in hybrid superconducting qubit architectures.

To determine the optimal waveform of the external flux bias for realizing fast, high-fidelity CZ gates, we employ a low-dimensional Fourier-cosine pulse parameterization together with a physically motivated cost function that independently penalizes conditional-phase errors and population leakage from the computational subspace. This approach substantially reduces the dimensionality of the optimization problem while preserving the dominant physical error mechanisms responsible for gate infidelity, thereby enabling efficient optimization of the coupler control waveform.

Numerical simulations based on the full system Hamiltonian demonstrate that the optimized control pulse realizes a microwave-free CZ gate within $25\,\mathrm{ns}$. The resulting gate exhibits an average fidelity exceeding $99.99\%$ together with negligible leakage, demonstrating that high-fidelity entangling operations can be achieved in hybrid superconducting architectures using only a single flux-control channel.

The remainder of this paper is organized as follows. In Sec.~II, we introduce the hybrid fluxonium-transmon architecture and present the theoretical model describing the flux-controlled tunable coupler. In Sec.~III, we elucidate the high-contrast ZZ coupling and the proposed CZ gate including the accumulation of the conditional phase, the Fourier-cosine pulse parameterization, and the optimization objective. In Sec.~IV, we present the numerical results, including the optimized control pulse, the population dynamics during the gate operation, and the resulting gate fidelity, followed by a discussion of the system decoherence on the gate performance and the spectator-induced gate error in Sec. V. Finally, the main conclusions and future perspectives are summarized in Sec.~VI.

\section{Model}

\begin{figure}
\begin{center}
\includegraphics[width = 8.50cm, height = 3.20cm]{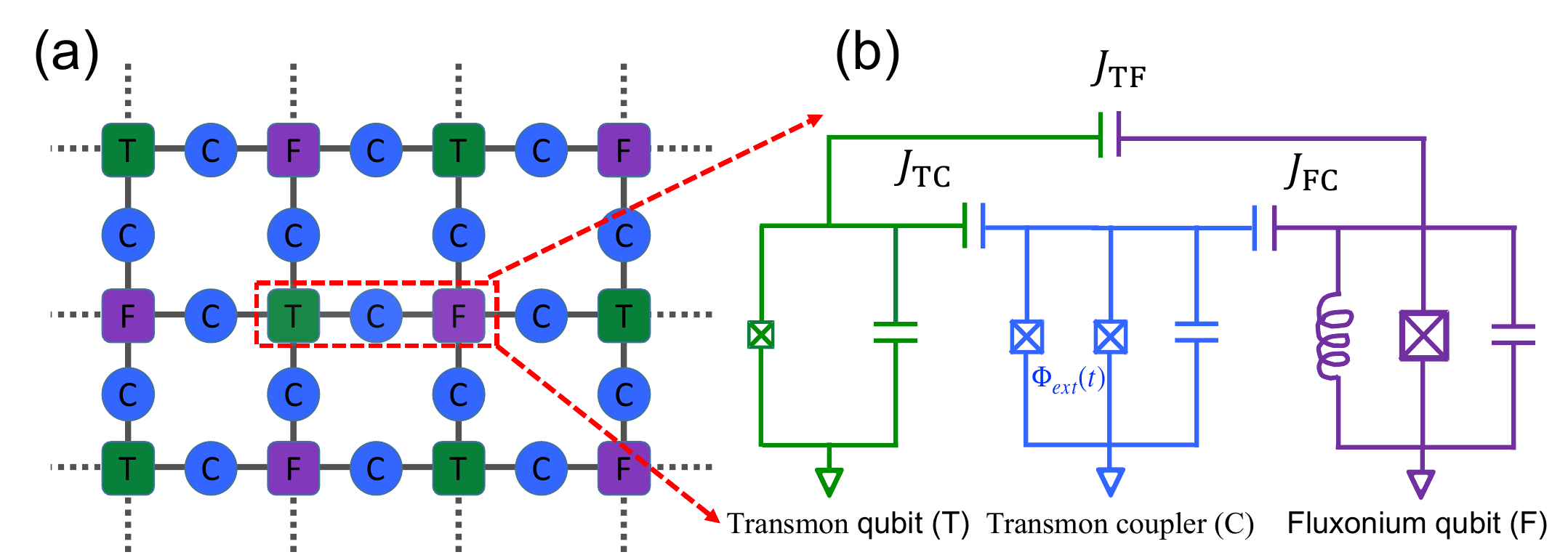}
\end{center}
\caption{(a) Schematic of a two-dimensional coupled lattice of a hybrid superconducting qubit system. Purple and green squares denote fluxonium (F) and transmon (T) qubits, respectively, and blue circles denote transmon couplers (C). Red dashed boxes indicate the smallest coupled model. (b) Schematic of the smallest coupled TCF circuit. The coupler mediates both a static direct qubit-qubit coupling and a dynamically tunable interaction controlled by the external flux $\Phi_{ext}(t)$.}
\label{Fig1ab}
\end{figure}

\subsection{Hybrid superconducting qubit architecture}

We consider a hybrid superconducting circuit composed of a fixed-frequency transmon qubit, a fluxonium qubit, and a flux-tunable transmon coupler, as schematically illustrated by the red-dotted box in Fig.~\ref{Fig1ab}(a). The transmon and fluxonium are capacitively coupled both directly and indirectly through the intermediate tunable coupler, as shown in Fig.~\ref{Fig1ab}(b).

This hybrid architecture combines the complementary advantages of the two superconducting qubit modalities. The transmon provides a simple control interface together with excellent coherence properties, whereas the fluxonium offers strong anharmonicity and enhanced robustness against charge noise. The tunable transmon coupler serves as a controllable quantum bus that mediates the effective interaction between the two qubits.

As in conventional tunable-coupler architectures, where the coupler serves as a dynamically tunable mediator of the effective qubit–qubit interaction~\cite{FeiYanPRApp100540622018, Mundada2019, HGotoPhysRevAppl180340382022, DLCampbellPRApp190640432023, BFoxenPRL1251205042020, YuanXuPhysRevLett1252405032020, JStehlikPhysRevLett1270805052021, YoungkyuSungPhysRevX110210582021, RuiLiPhysRevX140410502024}, the present scheme exploits the coupler as an active control element. A single external magnetic flux applied to the coupler continuously tunes its transition frequency and, consequently, modulates the effective interaction between the computational qubits. The entire gate operation can therefore be realized through a single time-dependent control parameter~\cite{YuanXuPhysRevLett1252405032020, JStehlikPhysRevLett1270805052021, RuiLiPhysRevX140410502024, MCCollodoPRL1252405022020}, substantially reducing the dimensionality of the control landscape while retaining sufficient control flexibility to engineer the interaction required for high-fidelity entangling gates.

The total Hamiltonian of the hybrid circuit can be written as

\begin{equation}
H(t)=H_{\mathrm T}+H_{\mathrm F}+H_{\mathrm C}(t)+H_{\mathrm int}(t),
\label{eq:Htotal}
\end{equation}
where $H_{\mathrm T}$, $H_{\mathrm F}$, and $H_{\mathrm C}(t)$ describe the transmon, the fluxonium, and the tunable coupler, respectively, while $H_{\mathrm int}(t)$ represents the interactions among them. The explicit time dependence originates solely from the external magnetic flux applied to the coupler.

\subsection{Hamiltonian of the hybrid circuit}

\subsubsection{Fixed-frequency transmon qubit}

The fixed-frequency transmon is described as a weakly anharmonic oscillator,
\begin{equation}
H_{\mathrm T} = \omega_{\mathrm T}\hat{a}^\dagger \hat{a} - \frac{\alpha_{\mathrm T}}{2}\hat{a}^\dagger \hat{a}^\dagger \hat{a}\hat{a},
\label{eq:HT}
\end{equation}
where $\hat{a} (\hat{a}^\dagger)$ denotes the annihilation (creation) operator. The transition frequency and anharmonicity are given by \cite{MKhezriThesis2018}
\begin{align}
\omega_{\mathrm T} &= \sqrt{8E_J^{(\mathrm T)}E_C^{(\mathrm T)}}\left(1-3\lambda_{\mathrm T}-9\lambda_{\mathrm T}^2\right),\nonumber\\
\alpha_{\mathrm T}  &= \sqrt{8E_J^{(\mathrm T)}E_C^{(\mathrm T)}}\left(3\lambda_{\mathrm T}+\frac{162}{8}\lambda_{\mathrm T}^2\right),
\end{align}
with
\begin{equation}
\lambda_{\mathrm T} = \frac13\sqrt{\frac{E_C^{(\mathrm T)}}{8E_J^{(\mathrm T)}}}.
\end{equation}
Here, $E_J^{(\rm{T})}$ and $E_C^{(\rm{T})}$ denotes the Josephson and charging energies of the transmon qubit. The corresponding charge operator is
\begin{equation}
\hat n_{\mathrm T} = \left(\frac{8E_C^{(\mathrm T)}}{E_J^{(\mathrm T)}}\right)^{-1/4}\frac{\hat{a}-\hat{a}^\dagger}{i\sqrt2}.
\end{equation}

\subsubsection{Fluxonium qubit}

The fluxonium Hamiltonian is
\begin{equation}
H_{\mathrm F} = 4E_C^{(\mathrm F)}\hat n^2 + \frac12E_L^{(\mathrm F)}\hat\phi^2 - E_J^{(\mathrm F)}\cos(\hat\phi-2\pi\Phi_{\mathrm F}),
\label{eq:HF}
\end{equation}
where $E_C^{(\mathrm F)}$, $E_L^{(\mathrm F)}$, and $E_J^{(\mathrm F)}$ denote the charging, inductive, and Josephson energies, respectively.
To facilitate numerical calculations, the Hamiltonian is expressed in the harmonic-oscillator basis,
\begin{align}
\hat\phi = \phi_0(\hat{b}+\hat{b}^\dagger), \hat n = n_0(\hat{b}^\dagger-\hat{b}),
\end{align}
with
\begin{align}
\phi_0 = \frac1{\sqrt2}\left(\frac{8E_C^{(\mathrm F)}}{E_L^{(\mathrm F)}}\right)^{1/4},
n_0 = \frac{i}{\sqrt2}\left(\frac{E_L^{(\mathrm F)}}{8E_C^{(\mathrm F)}}\right)^{1/4}.
\end{align}
The resulting Hamiltonian is numerically diagonalized, and only the lowest eigenstates are retained for subsequent simulations.

\subsubsection{Flux-tunable transmon coupler}

The central element of the present architecture is a flux-tunable transmon coupler. Its effective Josephson energy depends on the external magnetic flux,
\begin{equation}
E_J^{(\mathrm C)}(t) = E_{J,\max}\cos \left(\pi\Phi(t)\right),
\label{eq:EJflux}
\end{equation}
which leads to the effective Hamiltonian
\begin{equation}
H_{\mathrm C}(t) = \omega_{\mathrm C}(t)\hat{c}^\dagger \hat{c} - \frac{\alpha_{\mathrm C}(t)}2\hat{c}^\dagger \hat{c}^\dagger \hat{c}\hat{c}.
\end{equation}
The instantaneous transition frequency and anharmonicity are
\begin{align}
\omega_{\mathrm C}(t) &= \sqrt{8E_CE_J^{(\mathrm C)}(t)}\left(1-3\lambda(t)-9\lambda^2(t)\right),\nonumber\\
\alpha_{\mathrm C}(t) &=\sqrt{8E_CE_J^{(\mathrm C)}(t)}\left(3\lambda(t)+\frac{162}{8}\lambda^2(t)\right),
\end{align}
where
\begin{equation}
\lambda(t) = \frac13\sqrt{\frac{E_C}{8E_J^{(\mathrm C)}(t)}}.
\end{equation}

Similarly, the charge operator becomes
\begin{equation}
\hat n_{\mathrm C}(t) = \eta(t)(\hat{c}-\hat{c}^\dagger),
\end{equation}
with
\begin{equation}
\eta(t) = \left(\frac{8E_C}{E_J^{(\mathrm C)}(t)}\right)^{-1/4}\frac1{i\sqrt2}.
\end{equation}

\subsubsection{Interaction Hamiltonian}

The three subsystems interact through capacitive couplings,
\begin{equation}
H_{\mathrm int}(t) = H_{\mathrm{TF}} + H_{\mathrm {TC}}(t) + H_{\mathrm {FC}}(t),
\label{Hint}
\end{equation}
where
\begin{align}
H_{\mathrm {TF}} &= J_{\mathrm {TF}}\hat n_{\mathrm T}\hat n_{\mathrm F},\nonumber\\
H_{\mathrm {TC}}(t) &= g_{\mathrm {TC}}(t)\hat n_{\mathrm T}(\hat{c}-\hat{c}^\dagger),\nonumber\\
H_{\mathrm {FC}}(t) &= g_{\mathrm {FC}}(t)\hat n_{\mathrm F}(\hat{c}-\hat{c}^\dagger),
\end{align}
with $g_{\mathrm {TC}}(t) = J_{\mathrm {TC}}\eta(t)$ and $g_{\mathrm {FC}}(t) = J_{\mathrm {FC}}\eta(t)$.
Therefore, although the bare capacitive coupling constants remain fixed, the effective qubit-coupler interactions become continuously tunable through the external magnetic flux.

\subsection{Flux-controlled tunable coupler}

A key feature of the proposed scheme is that the entire gate operation is governed by a single external control parameter, namely the magnetic flux $\Phi_{ext}(t)$ applied to the coupler. Through Eq.~(\ref{eq:EJflux}), the flux modifies the effective Josephson energy of the coupler and consequently changes several physical quantities simultaneously,
\begin{equation}
\Phi(t)\longrightarrow E_J^{(\mathrm C)}(t)\longrightarrow\left\{\omega_{\mathrm C}(t),\alpha_{\mathrm C}(t),g_{\mathrm {TC}}(t),g_{\mathrm {FC}}(t)\right\}.
\end{equation}
Thus, one control waveform simultaneously determines the instantaneous spectrum of the coupler and its interactions with both qubits.

To accurately simulate the system dynamics while maintaining computational efficiency, each subsystem is truncated to its lowest relevant energy levels. The fixed-frequency transmon and the tunable coupler are each represented by their lowest three eigenstates, whereas the fluxonium qubit is truncated to its lowest five eigenstates after numerical diagonalization of Eq.~(\ref{eq:HF}). The resulting Hilbert space therefore has a dimension of $3 \times 5 \times 3 = 45$.

This truncation fully captures the dominant leakage channels relevant to the CZ gate operation while keeping the numerical simulations computationally tractable. Convergence with respect to the chosen Hilbert-space dimensions has been verified by increasing the truncation levels, and no observable changes in the optimized gate performance were found within numerical precision.

\section{Physical Mechanism of the CZ Gate}

\begin{figure}
\begin{center}
\includegraphics[width = 8.0cm, height = 6.0cm]{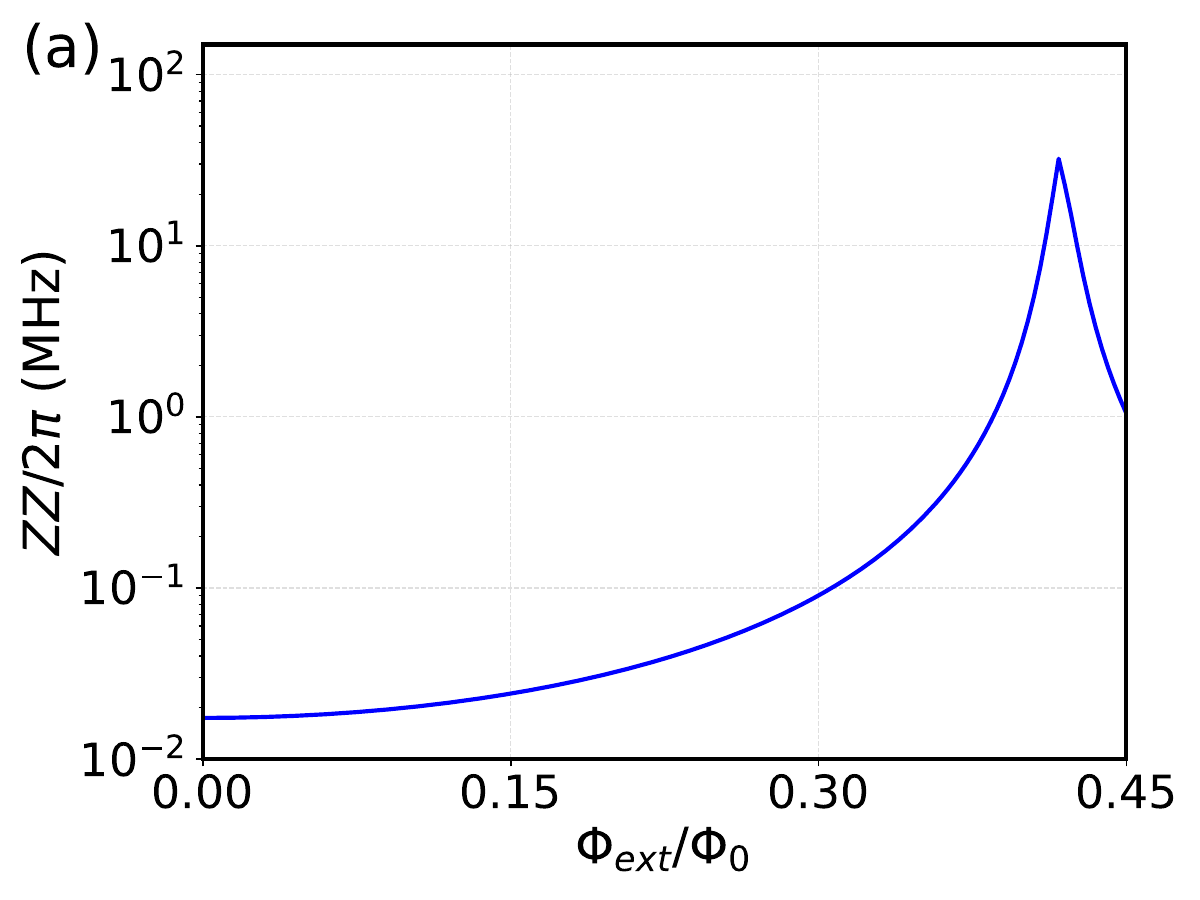}
\includegraphics[width = 8.0cm, height = 6.0cm]{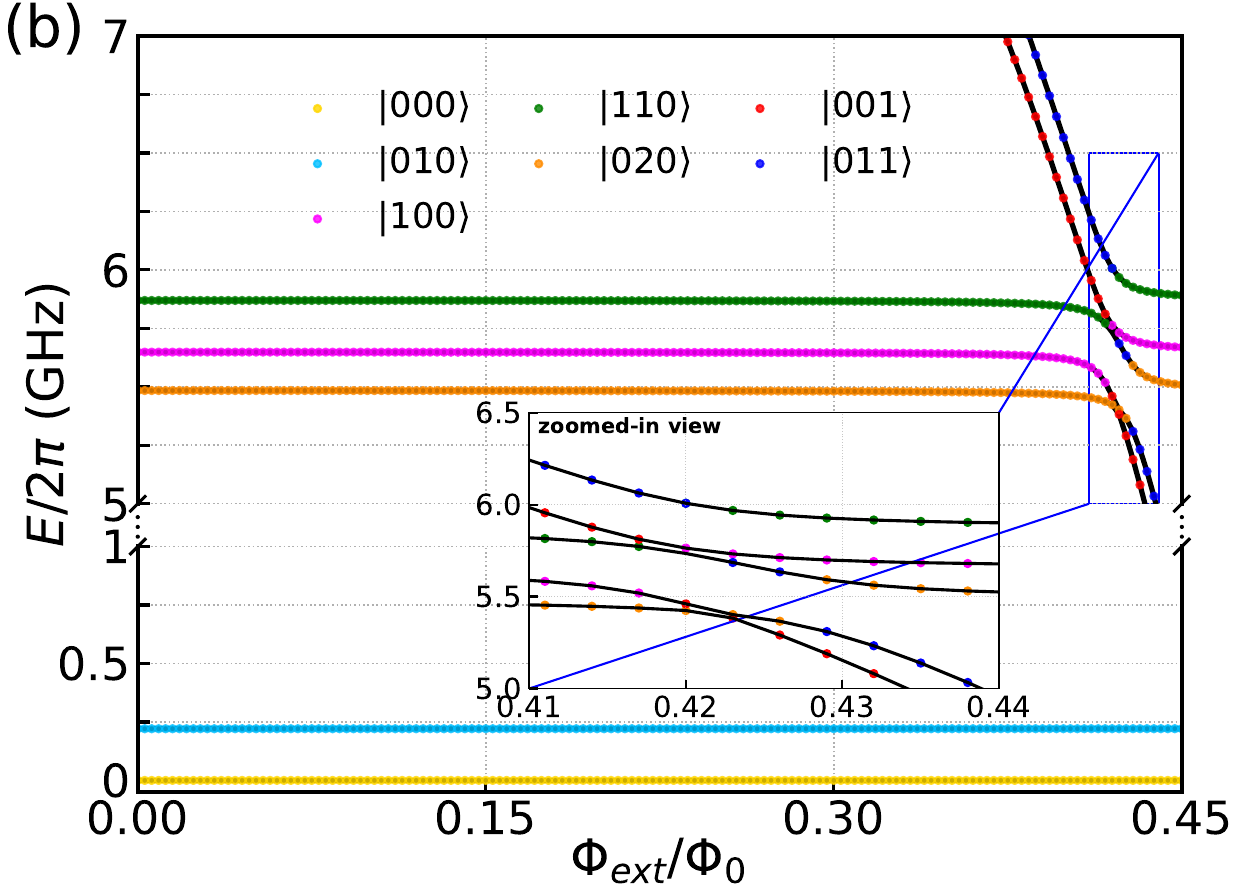}
\end{center}
\caption{(a) Calculated ZZ interaction strength between the fluxonium and transmon qubits as a function of the coupler flux bias $\Phi_{ext}(t)$. The coupler is assumed to remain in its ground state $\lvert 0\rangle$. The residual ZZ interaction is strongly suppressed when the coupler is biased near its maximum-frequency point, indicating that the effective qubit–qubit interaction can be substantially reduced through appropriate coupler biasing. (b) Energy spectrum of the hybrid superconducting architecture as a function of the external flux bias applied to the tunable coupler. The inset provides a magnified view of a representative avoided crossing.}
\label{Fig2ab}
\end{figure}

The realization of a high-fidelity controlled-Z gate in the proposed hybrid superconducting architecture relies on four essential ingredients. First, the residual static $\rm{ZZ}$ interaction is strongly suppressed by the tunable coupler, thereby minimizing unwanted conditional phase accumulation during idle operation. Second, the desired entangling phase is generated through the controlled evolution induced by the flux pulse. Third, the magnetic-flux waveform is parameterized using a smooth Fourier-cosine expansion that naturally satisfies the experimentally required boundary conditions. Finally, an optimization objective combining conditional-phase accuracy and leakage suppression is employed to determine the optimal pulse parameters.

\begin{table*}[t]
\centering
\caption{Hamiltonian parameters of the hybrid superconducting architecture shown in Fig. ~\ref{Fig1ab}(b) with spectator qubits}
\label{tab:hamiltonian}
\begin{tabular}{l
@{\hspace{18pt}}c
@{\hspace{18pt}}c
@{\hspace{18pt}}c
@{\hspace{18pt}}c
@{\hspace{18pt}}c}
\hline
\hline
Frequency (GHz) & $E_J/2\pi$ (GHz) & $E_C/2\pi$ (GHz) & $E_L/2\pi$ (GHz) & Coupling (MHz)  \\

\hline
Transmon $(\omega_T)$   
& 21.5   
& 0.20  
& --   
& $J_{TC}/2\pi=100$ \\

Fluxonium $(\omega_F)$  
& 5.71   
& 1.30  
& 0.59  
& $J_{TF}/2\pi=10$ \\

Coupler $(\omega_C)$        
& 55.0   
& 0.32  
& --   
& $J_{FC}/2\pi=250$ \\

\hline

Fluxonium Spectator $(\omega_{F_s})$        
& 5.7   
& 1.20  
& 0.65 
& $J_{TF_s}/2\pi=10$ \\

Transmon Spectator $(\omega_{T_s})$        
& 25.0   
& 0.20  
& --   
& $J_{FT_s}/2\pi=10$ \\

\hline
\hline
\label{Table1}
\end{tabular}
\end{table*}

\subsection{Suppression of residual ZZ interaction}
\label{sec:ZZ_suppression}

Residual $\rm{ZZ}$ interaction is an important source of coherent error in superconducting two-qubit gates. A finite static $\rm{ZZ}$ interaction produces a state-dependent frequency shift between the computational qubits and consequently leads to unwanted conditional-phase accumulation even in the absence of an intentional gate operation. Suppressing this residual interaction is therefore essential for minimizing coherent errors and improving the fidelity of entangling gates.

In the present hybrid TCF architecture, the residual $\rm{ZZ}$ interaction originates from multiple coupling pathways involving the fixed-frequency transmon, fluxonium, and tunable transmon coupler. In particular, the direct transmon--fluxonium coupling provides a finite background contribution to the $\rm{ZZ}$ interaction, while the coupler introduces additional flux-dependent interaction pathways. The interference among these pathways allows the effective $\rm{ZZ}$ interaction to be continuously modified by tuning the external magnetic flux applied to the coupler. A perturbative derivation of the corresponding analytical expression is provided in Appendix~\ref{app:analyticalZZ}.

For the parameters listed in Table~\ref{Table1}, we evaluate the residual $\rm{ZZ}$ interaction as a function of the external magnetic flux and compare the result obtained from the perturbative expression with that obtained from exact numerical diagonalization of the full multilevel Hamiltonian. The numerical $\rm{ZZ}$ interaction \cite{RJSchoelkopfNature4602009} is extracted from the dressed energies associated with the computational states $\ket{000}$, $\ket{010}$, $\ket{100}$, and $\ket{110}$ according to
\begin{equation}
\rm{ZZ}_{\rm num} = (E_{110}-E_{100})-(E_{010}-E_{000}).
\label{eq:ZZ_numerical}
\end{equation}
Here, the basis states are defined in the order $\ket{\rm{TFC}}$, where the first, second, and third indices denote the transmon, fluxonium, and coupler states, respectively, and the coupler is assumed to remain in its ground state $\ket{0}$. When confined to the qubit subspace, we use the notation $\ket{\rm{TF}}$. 

The numerical results is shown in Fig.~\ref{Fig2ab}(a). An important feature of the calculated $\rm{ZZ}$ interaction is its strong suppression when the coupler is biased near its maximum-frequency point. At this operating point, the coupler is far detuned from the computational qubits, thereby reducing the strength of the coupler-mediated interaction. More importantly, the remaining direct and coupler-mediated contributions can partially cancel each other through destructive interference, see Appendix~\ref{app:analyticalZZ} for more details. Consequently, the effective $\rm{ZZ}$ interaction can be reduced to a very small value at the idle configuration, where the coupler is biased near its maximum frequency point $(\Phi_{ext}=0)$.

The energy-level structure provides a complementary picture of this suppression mechanism. Using the Hamiltonian parameters listed in Table~\ref{Table1}, we calculate the dressed energy spectrum of the hybrid TCF architecture as a function of the external magnetic flux applied to the tunable coupler, as shown in Fig.~\ref{Fig2ab}(b). The computational states $\ket{000}$, $\ket{010}$, $\ket{100}$, and $\ket{110}$ are identified according to their dominant bare-state compositions~\cite{AGaliautdinovPRA850423212012}. As the external flux is varied, the coupler transition frequency changes accordingly, causing the computational states to approach and hybridize with nearby noncomputational states. This hybridization manifests itself as a series of avoided crossings in the energy spectrum.

The avoided crossings are particularly important for understanding both the origin of the effective interaction and the leakage mechanisms during the CZ operation. At an avoided crossing, the coupling between two bare states produces level repulsion and modifies the corresponding dressed-state energies. Since the conditional interaction is determined by the relative energy shifts of the four computational states, these coupling-induced level shifts directly contribute to the effective $\rm{ZZ}$ interaction. The inset of Fig.~\ref{Fig2ab}(b) provides a magnified view of a representative avoided crossing, clearly illustrating the local hybridization of the energy levels induced by the coupler-mediated interaction.

The same energy-level structure also determines the dominant leakage channels during the gate operation. By examining the neighboring noncomputational states involved in the relevant avoided crossings, one can identify the states that are most strongly coupled to each computational state. The optimized flux pulse must therefore balance two competing requirements: it should bring the system sufficiently close to the relevant interaction region to accumulate the desired conditional phase, while avoiding excessive population transfer to noncomputational states. This provides the physical basis for the pulse-optimization strategy employed in the CZ-gate protocol.

Along the optimized control trajectory $\Phi_{\rm ext}(t)$, the coupler therefore plays a dual role. On the one hand, its flux-dependent transition frequency provides a controllable interaction pathway that enables the generation of the desired conditional phase. On the other hand, appropriate biasing of the coupler allows the unwanted static $\rm{ZZ}$ interaction to be strongly suppressed through destructive interference among different virtual transition pathways.

\subsection{Conditional phase accumulation}

Once the residual ZZ interaction at the idle configuration has been minimized, the entangling operation is generated through controlled dynamical phase accumulation.

Within the computational basis $\{|00\rangle,|01\rangle,|10\rangle,|11\rangle\}$, the evolution operator after the flux pulse can always be written as $U = \mathrm{diag}\left(e^{i\phi_{00}}, e^{i\phi_{01}}, e^{i\phi_{10}}, e^{i\phi_{11}}\right)$, up to leakage outside the computational subspace.

The physically relevant quantity is not the individual phases but the conditional phase $\phi_{\rm cond} = \phi_{11} - \phi_{10} - \phi_{01} + \phi_{00}$, which is invariant under both global phases and local single-qubit Z rotations. Therefore, it uniquely characterizes the nonlocal entangling interaction. An ideal CZ gate is obtained whenever $\phi_{\rm cond} = \pi \quad (\mathrm{mod}\;2\pi)$.

The optimization therefore aims to engineer the flux pulse such that the desired conditional phase is accumulated while simultaneously suppressing leakage into higher excited states.

\subsection{Fourier-cosine flux pulse parameterization}

To generate experimentally feasible control pulses, the external magnetic flux is parameterized using a truncated Fourier-cosine expansion. The auxiliary pulse function is written as~\cite{JMMartinisPRA900223072014}
\begin{equation}
P(t) = \sum_{k=1}^{N}c_k\left[1-\cos\left(\frac{2k\pi t}{t_f}\right)\right],
\label{eq:fourierpulse}
\end{equation}
with $t_f$ denoting the gate length. Unlike arbitrary polynomial parameterizations, the Fourier-cosine basis naturally satisfies $P(0)=0, P(t_f)=0$, together with $\dot P(0)= \dot P(t_f)=0$. These boundary conditions ensure that the magnetic flux and its first derivative vary smoothly throughout the gate operation, thereby reducing high-frequency spectral components that could induce unwanted nonadiabatic excitations.

The pulse function determines an interpolation angle
\begin{equation}
\theta(t) = \theta_i + \frac{\theta_f-\theta_i}{2}P(t),
\end{equation}
where $\theta_i = \tan^{-1}[\frac{2g_{\rm{TF}}}{\omega_C(0) - \omega_{\mathrm T} }]$, from which the instantaneous coupler frequency is obtained,
\begin{equation}
\omega_{\mathrm C}(t) = \omega_{\mathrm T} + \frac{2g_{\rm{TF}}}{\tan\theta(t)}.
\end{equation}
Here, $g_{\rm{TF}}$ denots the strength for the interaction between the transmon qubit and the coupler at the on-resonance point, see Fig.~\ref{Fig2ab}(b). Finally, the required magnetic flux is obtained through the inverse mapping
\begin{equation}
\Phi_{ext}(t) = f^{-1}\left(\omega_{\mathrm C}(t)\right),
\end{equation}
where the function $f(\Phi)$ is determined numerically from the coupler spectrum.

In the present work, only three Fourier coefficients are employed, corresponding to a low-dimensional optimization problem while still providing sufficient flexibility for realizing high-fidelity gate operations.

\subsection{Optimization objective}

The objective of the optimization is to determine the Fourier-cosine pulse parameters that realize an ideal controlled-phase gate while minimizing unwanted leakage. Instead of directly maximizing the average gate fidelity, we optimize two physically independent quantities: (1) the conditional phase error, (2) the leakage probability outside the computational subspace. The conditional phase error is defined as
phase error,
\begin{equation}
\delta_\phi = |\phi_{\rm{cond}}|-\pi,
\end{equation}
which measures the deviation from the ideal CZ phase.
The leakage error is evaluated through the projector onto the computational subspace, can be defined as~\cite{CJWoodPRA970323062018, PXuPRA1080326152023}
\begin{equation} 
\begin{aligned} 
L_e = 1 - \frac{\sum_{m, n \in \{0,1\}}\text{Tr}[ U^{\dagger}_{\text{real}}|mn\rangle\langle mn|U_{\text{real}}]}{4}, 
\end{aligned} 
\end{equation}
where $U_{\text{real}}$ represents the actual evolution within the computational subspace, derived from the full system Hamiltonian, excluding decoherence effects.
The optimization minimizes the composite cost function
\begin{equation}
C = (\delta_\phi)^2 + L_e.
\label{eq:cost}
\end{equation}
This choice has a clear physical interpretation. The first term ensures that the accumulated nonlocal phase approaches the ideal value of $\pi$, thereby determining the entangling capability of the gate. The second term suppresses population transfer into non-computational states.

\section{Numerical Results}

In this section, we present the numerical results obtained from the optimal-control procedure described above. We first introduce the optimized magnetic-flux pulse obtained from the optimization. Then verify its performance by examining the population dynamics during the gate operation, followed by a quantitative evaluation of the average gate fidelity. Finally, we discuss the physical origin of the high gate performance and the advantages of the proposed control protocol.

\subsection{Optimized Flux Pulse}

\begin{figure}
\begin{center}
\includegraphics[width = 6.50cm, height = 5.00cm]{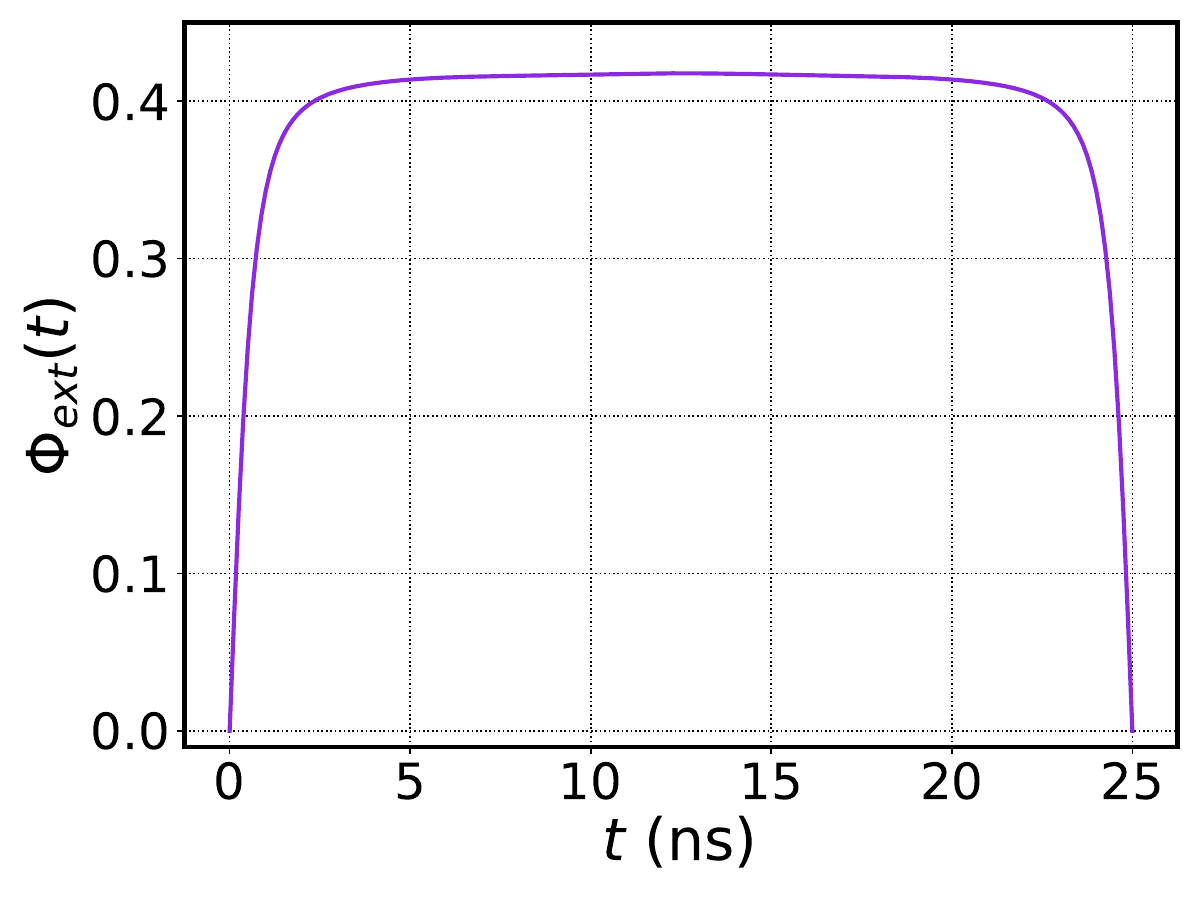}
\end{center}
\caption{Typical optimized magnetic-flux pulse applied to the tunable coupling transmon for $N=3$ with 25 ns. The pulse is parameterized by the optimized Fourier-cosine coefficients.}
\label{Fig3OptimalPulse}
\end{figure}

By minimizing the cost function defined in Eq.~(\ref{eq:cost}), we obtain an optimized magnetic-flux pulse capable of implementing a high-fidelity controlled-$\rm{Z}$ gate. For the Fourier-cosine parameterization with three basis functions ($N=3$) [giving rise to totally three free parameters ($c_1$, $c_2$, $\theta_f$) to be optimized], the typical optimized flux waveform is shown in Fig.~\ref{Fig3OptimalPulse}. Owing to the Fourier-cosine parameterization, the pulse varies smoothly throughout the gate operation while automatically satisfying the required boundary conditions at both the beginning and the end of the control sequence. Such smooth waveforms are advantageous for experimental implementation because they suppress high-frequency spectral components and reduce waveform distortion caused by finite-bandwidth control electronics.

\begin{figure*}
\begin{center}
\includegraphics[width = 17.50cm, height = 5.00cm]{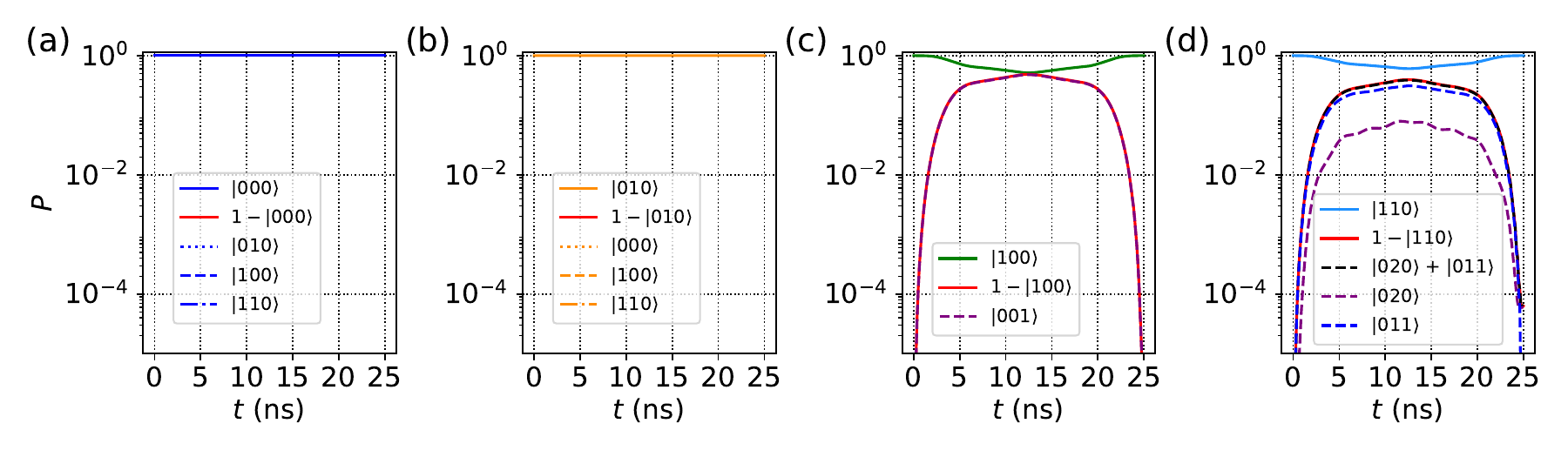}
\end{center}
\caption{Time evolution of the computational-state populations and leakage populations during the implementation of the CZ gate under the optimized magnetic-flux pulse. Panels (a)-(d) correspond to the initial states $|00\rangle$($|000\rangle$), $|01\rangle$($|010\rangle$), $|10\rangle$($|100\rangle$), and $|11\rangle$($|110\rangle$), respectively. The applied system parameters are outlined in Table~\ref{Table1} and optimized flux pulse is shown in Fig.~\ref{Fig3OptimalPulse}. }
\label{Fig4Dynamics}
\end{figure*}

For a typical optimized pulse shown in Fig.~\ref{Fig3OptimalPulse}, the phase error is below $6\times10^{-5}$ rad, and the population leakage remains below $1\times10^{-5}$. This demonstrates that the proposed scheme achieves high-fidelity conditional-phase accumulation with negligible leakage from the computational subspace.

\subsection{Population Dynamics}

To validate the optimized pulse beyond the final optimization metrics, we simulate the complete time evolution governed by the time-dependent Hamiltonian. Figure~\ref{Fig4Dynamics} presents the population dynamics for the four computational basis states, $|000\rangle$, $|010\rangle$, $|100\rangle$, $|110\rangle$, which are separately chosen as the initial states.

For the initial states $|000\rangle$ and $|010\rangle$, the evolution remains almost entirely inside the computational subspace throughout the gate operation. At the completion of the CZ gate, the leakage population from the computational subspace remains below $10^{-4}$. For the initial states $|100\rangle$ and $|110\rangle$, the optimized pulse intentionally induces temporary occupation of higher excited states. This transient population transfer is a characteristic feature of nonadiabatic entangling gates and provides the virtual excitation pathways required for accumulating the desired conditional phase. Importantly, the leaked population is coherently transferred back into the computational subspace before the gate terminates, leaving a residual leakage probability below $10^{-5}$.

These results show that the optimized pulse effectively utilizes intermediate excited states as virtual resources for conditional-phase accumulation while suppressing final leakage, thereby preserving the logical quantum information within the computational subspace. Furthermore, stronger coupling enables a substantial reduction in the gate duration. The corresponding performance in the strong-coupling regime is analyzed in detail in Appendix~\ref{app:strongcoupling}.

\subsection{Gate Fidelity}

Although the optimization is performed using the physically motivated cost function rather than the average gate fidelity itself, the final gate performance is evaluated using the standard state-averaged fidelity, which provides a rigorous and widely accepted benchmark for quantum gate operations.

Following Refs.~\cite{KlausPLA367472007}, the average gate fidelity is defined as
\begin{equation}
F_{\rm avg} = \frac{\mathrm{Tr}\left(U_{\rm real}^{\dagger}U_{\rm real}\right) + \left|\mathrm{Tr}\left(U_{\rm ideal}^{\dagger}U_{\rm real}\right)\right|^2}{20},
\end{equation}
where $U_{\rm real}$ is the projected evolution operator in the computational subspace and $U_{\rm ideal}$ denotes the ideal CZ gate.

During the gate operation, additional single-qubit dynamical phases are accumulated together with the desired nonlocal conditional phase. Since these local phases can be removed by virtual $\rm{Z}$ gates without increasing either the gate duration or the experimental complexity~\cite{JMGambettaPRA960223302017}, they are compensated before evaluating the gate fidelity. Using the optimized control pulse, we obtain an average gate fidelity exceeding $F_{\rm avg}>99.99\%$, while maintaining a leakage probability below $10^{-5}$.

The simultaneously high fidelity and extremely low leakage confirm that the proposed optimization strategy accurately captures the dominant physical error mechanisms governing the gate performance. Furthermore, the excellent agreement between the optimization objective and the final gate fidelity is consistent with the perturbative analysis, where the adopted cost function is shown to represent the second-order approximation of the average gate infidelity~\cite{QFicheuxPRX110210262021}.

\section{Discussion}

The above results establish the effectiveness of the proposed optimization scheme for realizing high-fidelity CZ gates in the hybrid superconducting architecture. We next examine the robustness of the optimized gate from two complementary perspectives. First, we study the dependence of the gate fidelity on the gate duration and investigate the impact of realistic decoherence processes, thereby characterizing the trade-off between gate speed and control accuracy. Second, we investigate the effect of spectator qubits in an extended coupled architecture, with particular attention to spectator-state-dependent crosstalk. These analyses provide further insight into the practical performance and scalability of the proposed CZ-gate scheme.

\subsection{Different gate times and effect of decoherence}

\begin{figure}
\begin{center}
\includegraphics[width = 7.50cm, height = 5.50cm]{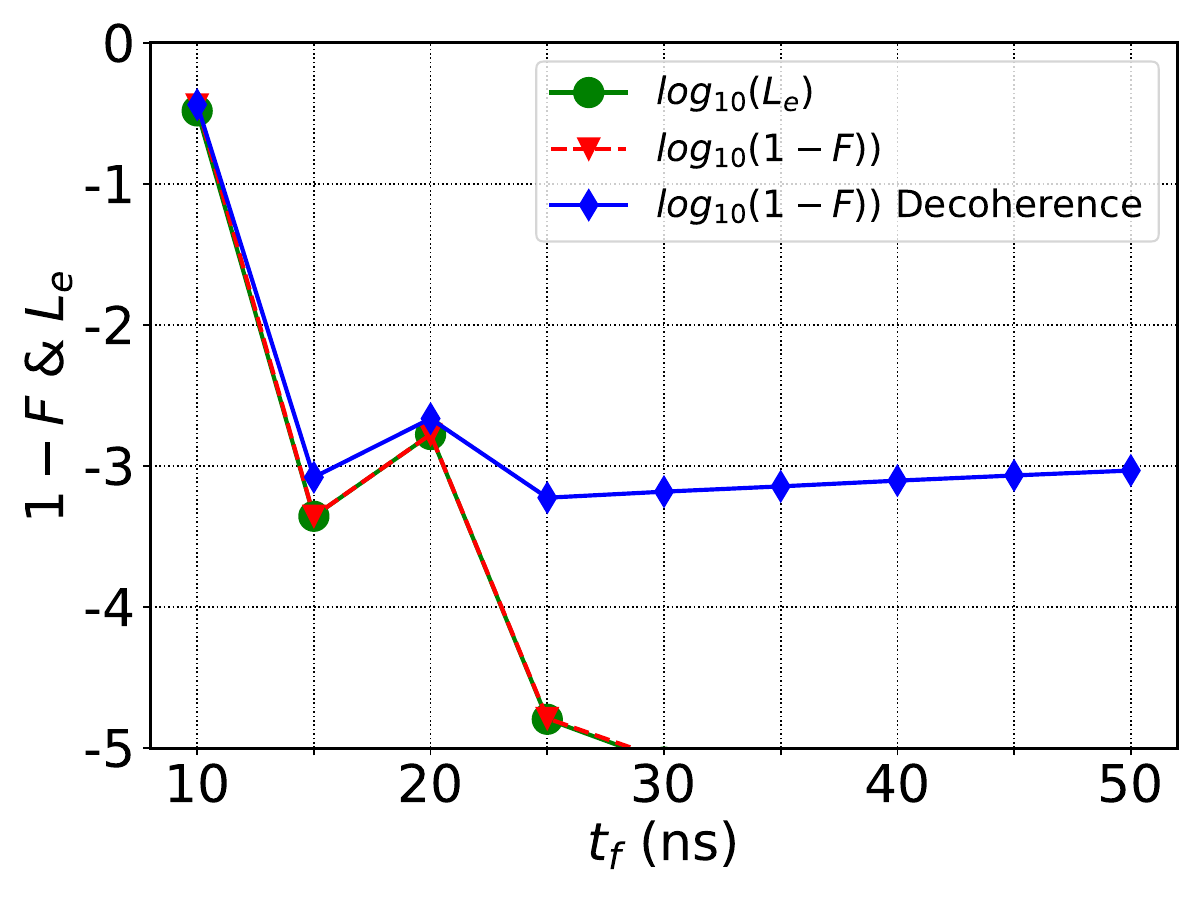}
\end{center}
\caption{Gate performance of the optimized CZ gate as a function of the gate duration $t_f$. The gate infidelity $(1-F)$ and leakage error $L_e$ are shown for the proposed configuration, with the decoherence-limited results obtained from the Lindblad master equation.The gate duration is varied from $10$ to $50$~ns.}
\label{Fig5}
\end{figure}

The results presented above focus primarily on a gate time of $T_{\rm g}=25$~ns. To further investigate the performance and applicability of the proposed optimization scheme over a broader time scale, we systematically optimize the control-flux pulse for different gate durations, ranging from 10 to 60~ns with a 5~ns increment. The optimized CZ-gate fidelity as a function of the gate time is shown in Fig.~\ref{Fig5}.

For each gate duration, the control-flux waveform is independently optimized to maximize the CZ-gate fidelity while suppressing leakage out of the computational subspace. In the absence of decoherence, the proposed optimization scheme achieves a fidelity exceeding $99.999\%$ for gate times longer than approximately 25~ns. This result demonstrates that the high-fidelity performance is not restricted to a single operating point, but can be maintained over a relatively broad range of gate durations. The dependence of the fidelity on the gate time originates from the competition between nonadiabatic transitions, leakage, and the accumulated dynamical phase during the flux-controlled evolution.

To further assess the robustness of the optimized gate under realistic conditions, we incorporate energy relaxation and pure dephasing through the Lindblad master equation \cite{LMEHJCarmichael2002},
\begin{equation}
\dot{\rho} = -i[H(t),\rho] + \sum_k \mathcal{D}[L_k]\rho,
\end{equation}
where
\begin{equation}
\mathcal{D}[L_k]\rho = L_k\rho L_k^\dagger -\frac{1}{2} \left\{L_k^\dagger L_k,\rho \right\},
\end{equation}
and $L_k$ denotes the collapse operator associated with the 
corresponding decoherence channel.

For the transmon qubit and the tunable transmon coupler, the relaxation and pure-dephasing collapse operators are given by
\begin{equation}
L_{\rm{T},1}=\sqrt{\frac{1}{T_{1,\rm{T}}}}\,\hat{a}, \qquad L_{\rm{T},\phi} = \sqrt{\frac{2}{T_{\phi,\rm{T}}}}\, \hat{a}^\dagger \hat{a},
\end{equation}
and
\begin{equation}
L_{\rm{C},1}=\sqrt{\frac{1}{T_{1,\rm{C}}}}\,\hat{c}, \qquad L_{C,\phi} = \sqrt{\frac{2}{T_{\phi,\rm{C}}}}\, \hat{c}^\dagger \hat{c},
\end{equation}
respectively. We take
$T_{1,\rm{T}}=100~\mu{\rm s}$ and $T_{\phi,\rm{T}}=200~\mu{\rm s}$ for the transmon, and $T_{1,\rm{C}}=20~\mu{\rm s}$ and $T_{\phi,\rm{C}}=40~\mu{\rm s}$ for the coupler~\cite{SiddiqiNatRevMater68752021}.

For the fluxonium qubit, we explicitly account for the level-dependent relaxation and pure-dephasing processes. The relaxation channels considered in the simulation are $|1\rangle\rightarrow|0\rangle$, $|2\rangle\rightarrow|1\rangle$, $|3\rangle\rightarrow|0\rangle$, and $|4\rangle\rightarrow|1\rangle$, with the corresponding collapse operators
\begin{equation}
\begin{aligned}
L_{\rm{F},10}&= \sqrt{\frac{1}{T_{1}^{10}}} |0\rangle\langle1|,L_{\rm{F},21}&= \sqrt{\frac{1}{T_{1}^{21}}} |1\rangle\langle2|,\\
L_{\rm{F},30}&= \sqrt{\frac{1}{T_{1}^{30}}} |0\rangle\langle3|,L_{\rm{F},41}&= \sqrt{\frac{1}{T_{1}^{41}}} |1\rangle\langle4|.
\end{aligned}
\end{equation}
The corresponding relaxation times are~\cite{Nguyen2022, ASomoroffPhysRevLett1302670012023}
\begin{equation}
T_{1}^{10}=100~\mu{\rm s}, \qquad
T_{1}^{21}=T_{1}^{30}=T_{1}^{41}=10~\mu{\rm s}.
\end{equation}

The level-dependent pure-dephasing channels are described by
\begin{equation}
L_{\rm{F},\phi}^{(m)} = \sqrt{\frac{2}{T_{\phi}^{(m)}}} |m\rangle\langle m|,\qquad m=1,2,3,4,
\end{equation}
with~\cite{Nguyen2022, ASomoroffPhysRevLett1302670012023}
\begin{equation}
T_{\phi}^{(1)}=200~\mu{\rm s}, \qquad T_{\phi}^{(2)} =T_{\phi}^{(3)}=T_{\phi}^{(4)}=20~\mu{\rm s}.
\end{equation}
Thus, the decoherence model explicitly includes both energy relaxation and level-dependent pure dephasing for the hybrid two-qubit system.

According to the fidelity definition with our previous work ~\cite{PengZhaoQST2025}, as shown in Fig.~\ref{Fig5}, the inclusion of decoherence naturally reduces the gate fidelity, particularly for longer gate durations. Nevertheless, the optimized CZ gate maintains a fidelity above $99.9\%$ over a gate-time range of approximately 25--50~ns, even in the presence of finite relaxation and dephasing times. This result indicates that the proposed control protocol possesses a considerable margin against decoherence and does not require an excessively short gate duration to achieve high-fidelity operation. At sufficiently long gate times, however, the benefit of further suppressing nonadiabatic leakage is gradually offset by the increased exposure to decoherence. We note that the decoherence rates used in the above analysis can generally be obtained reproducibly in current quantum systems, and can therefore be interpreted as a conservative estimate. Consequently, even high gate fidelity should be achievable upon further improvement of the system coherence times.

\subsection{Impact of Spectator Qubits on Gate Performance}
\label{sec:spectator}

\begin{figure}
\begin{center}
\includegraphics[width = 8.00cm, height = 1.50cm]{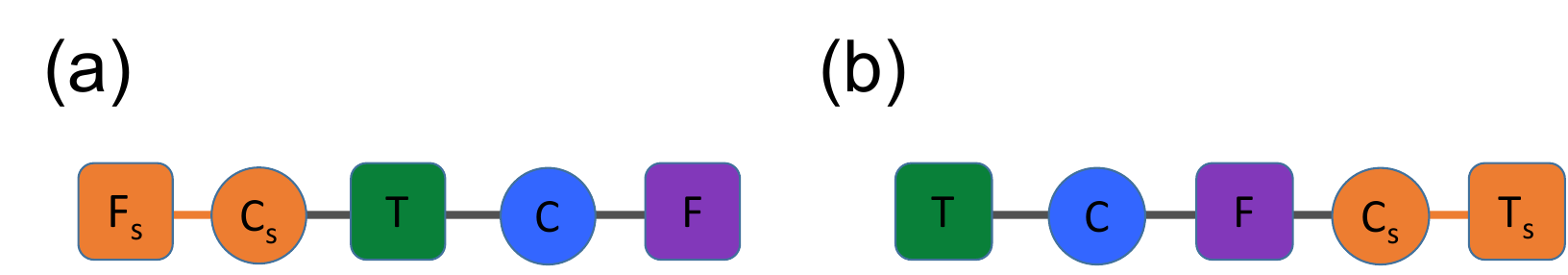}
\includegraphics[width = 8.00cm, height = 6.0cm]{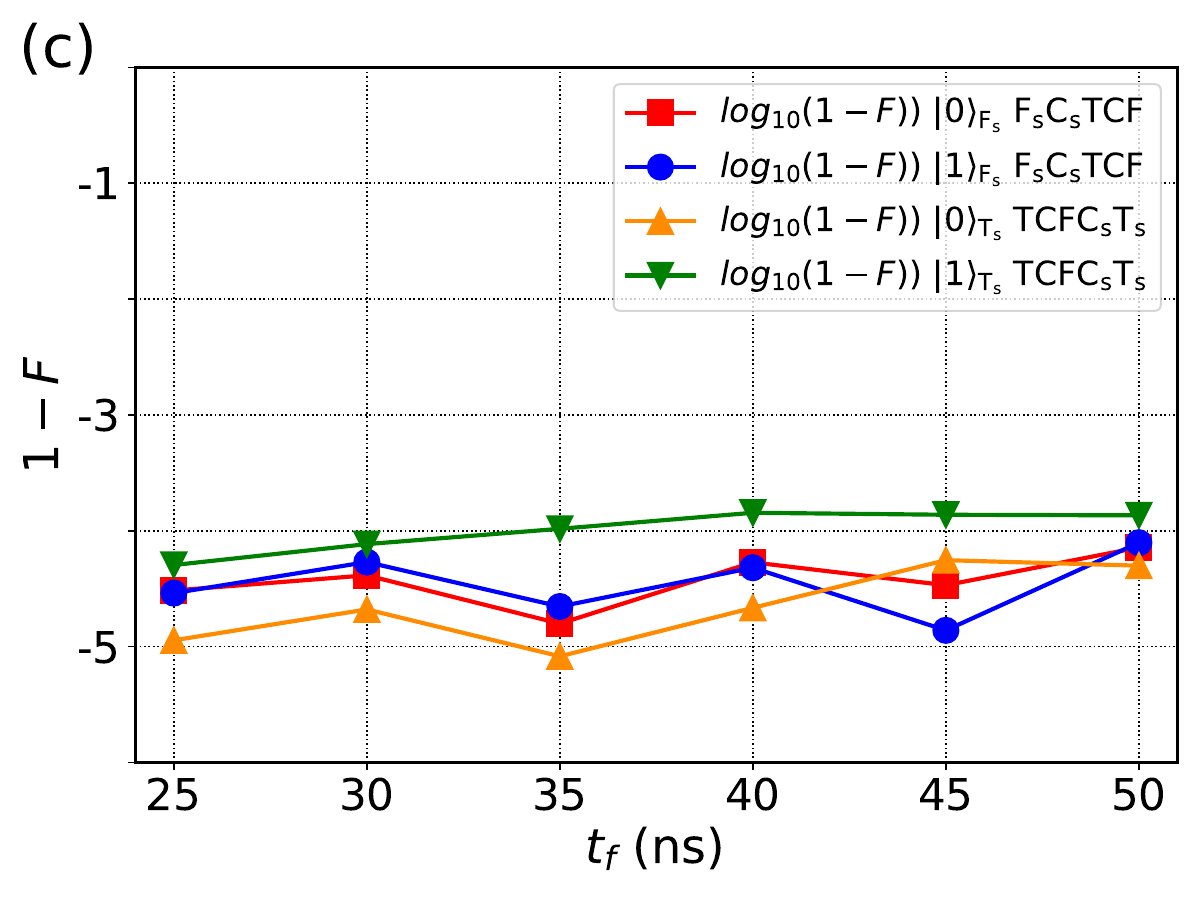}
\end{center}
\caption{Extended superconducting architectures with spectator qubits. (a) $\rm{F_sC_sTCF}$ configuration with a spectator fluxonium qubit $F_s$. (b) $\rm{TCFC_sT_s}$ configuration with a spectator transmon qubit $\rm{T_s}$. In both configurations, $\rm{C_s}$ denotes the spectator coupler connecting the spectator subsystem to the $\rm{TCF}$ system. (c) Gate infidelity, $(1-F)$, versus gate duration for the extended $\rm{F_sC_sTCF}$ and $\rm{TCFC_sT_s}$ architectures with spectator qubits. Results are shown for $t_f=25$, $30$, $35$, $40$, $45$, and $50$~ns, with the spectator qubit initialized in $\ket{0}$ or $\ket{1}$ and the spectator coupler fixed in $\ket{0}$. The low infidelity maintained over the entire gate-time range demonstrates the robustness of the optimized CZ gate against spectator-induced interactions.}
\label{Fig6}
\end{figure}

To further assess the scalability of the proposed CZ-gate scheme, we extend the three-element $\rm{TCF}$ subsystem to larger coupled architectures, as illustrated by the two-dimensional coupling lattice in Fig.~\ref{Fig1ab}(a). Specifically, we consider two extended configurations, $\rm{F_sC_sTCF}$ and $\rm{TCFC_sT_s}$, corresponding to Figs.~\ref{Fig6}(a) and \ref{Fig6}(b), respectively. Here, $\rm{F_s}$ ($\rm{T_s}$) denotes a spectator fluxonium (transmon) qubit, while $\rm{C_s}$ denotes a spectator transmon coupler with the same parameters as the tunable coupler $\rm{C}$ in the original $\rm{TCF}$ subsystem. The spectator coupler is maintained in its ground state $\ket{0}$, whereas the spectator qubit is initialized in either $\ket{0}$ or $\ket{1}$. To evaluate the robustness of the proposed control protocol in the presence of neighboring qubits, we first optimize the control pulse with the spectator qubit initialized in $\ket{0}$ and then keep the resulting optimized pulse parameters fixed when evaluating the gate fidelity for the spectator qubit in $\ket{1}$~\cite{SKrinnerPRApp140240422020, TQCaiPRL1270605052021, PengZhaoPRApp200540332023}. In this way, the comparison between the two spectator states directly characterizes the sensitivity of the optimized gate to the quantum state of the neighboring spectator qubit, rather than reflecting independent pulse optimization for each spectator state. Using the system parameters listed in Table~\ref{Table1}, we apply the same pulse-optimization procedure developed for the original $\rm{TCF}$ subsystem.

For the extended configuration containing a spectator fluxonium qubit, $\rm{F_sC_sTCF}$~\cite{SKrinnerPRApp140240422020, TQCaiPRL1270605052021, PengZhaoPRApp200540332023}, the optimized CZ gate remains highly robust against the additional degrees of freedom introduced by the neighboring subsystem. When the spectator fluxonium is initialized in $\ket{0}$ and $\ket{1}$, respectively, the resulting CZ-gate fidelities exceed $99.9969\%$ and $99.9971\%$, indicating a negligible spectator-induced correclated error (which is particuly hamful for quantum error correction).

We further examine the dependence of the gate performance on the gate duration. Figure~\ref{Fig6}(c) summarizes the gate infidelity, represented by $(1-F)$, for gate durations ranging from $25$ to $50$~ns. The six considered gate times, $t_g=25$, $30$, $35$, $40$, $45$, and $50$~ns, allow us to assess whether the robustness against spectator qubits is confined to a particular operating point or persists over a broader temporal regime. For the $\rm{F_sC_sTCF}$ configuration, both spectator states exhibit consistently low infidelities throughout the investigated time window. In particular, the infidelity for the $\ket{1}$ spectator state varies relatively smoothly with the gate duration, while that for the $\ket{0}$ state remains at a comparable level. These results indicate that the perturbation introduced by the spectator fluxonium is effectively controlled over the entire range of gate durations considered.

A similar analysis is performed for the $\rm{TCFC_sT_s}$ configuration containing a spectator transmon. The gate infidelity remains at or below the $10^{-4}$ level throughout the $25$--$50$~ns range for both spectator states. Although a moderate dependence on the gate duration is observed, the overall variations remain small compared with the high fidelity maintained throughout the investigated regime. The spectator transmon initialized in $\ket{1}$ generally exhibits somewhat larger fluctuations in the gate infidelity than the $\ket{0}$ case, reflecting its stronger participation in the coupled energy-level structure. Nevertheless, even for the excited spectator state, the optimized gate maintains a fidelity above $99.99\%$ over the entire gate-time range considered.

A comparison of the four cases further shows that the spectator-induced error does not increase monotonically with the gate duration. Instead, because the control pulse is reoptimized for each gate duration, the corresponding flux trajectory adapts to the available evolution time, resulting in a nonmonotonic but well-confined variation of the residual gate error. This behavior demonstrates that the optimization procedure can accommodate the additional interaction pathways introduced by neighboring qubits over a broad range of gate durations, rather than being restricted to a single isolated operating point.

Taken together, the results over the $25$--$50$~ns gate-time range demonstrate that the proposed optimization protocol is robust not only to the type and initial state of the spectator qubit, but also to variations in the gate duration. The spectator-induced fidelity degradation remains below $10^{-4}$ across the investigated extended configurations and temporal range, providing strong evidence that the optimized control strategy can tolerate additional coupled degrees of freedom. This robustness is particularly important for the two-dimensional coupled architecture shown in Fig.~\ref{Fig1ab}(a), where multiple neighboring qubits and couplers can introduce additional energy levels and interaction pathways. Therefore, these results support the scalability of the proposed CZ-gate protocol toward larger hybrid superconducting quantum architectures.


\section{Conclusion}

In this work, we have proposed a high-fidelity controlled-Z (CZ) gate for a hybrid superconducting architecture comprising a fluxonium qubit, a fixed-frequency transmon qubit, and a flux-tunable transmon coupler. The gate is realized solely through the temporal modulation of a single external magnetic-flux control pulse applied to the coupler, which simultaneously tunes its transition frequency and the effective qubit-qubit interactions. Based on the full system Hamiltonian, we show that the residual static ZZ interaction is strongly suppressed at the idle point, thereby reducing unwanted conditional-phase accumulation and spectator-induced coherent errors. To realize the desired CZ operation, we employ a smooth Fourier-cosine pulse parameterization that satisfies the required boundary conditions and avoids abrupt control variations. A physically motivated cost function comprising the conditional-phase error and leakage from the computational subspace is then adopted to separately suppress the dominant error mechanisms, while maintaining a direct connection to the average gate infidelity in the perturbative regime.

The optimized pulse produces a conditional phase close to the ideal value of $\pi$ while keeping the final leakage population below $10^{-5}$. Although transient population transfer to auxiliary states occurs during the gate operation, the population returns almost completely to the computational subspace at the end of the pulse. The resulting average gate fidelity exceeds $99.99\%$, demonstrating that the proposed protocol combines fast operation, effective leakage suppression, and high-fidelity gate performance. These results establish single-parameter flux control as a simple and robust approach for implementing high-fidelity entangling gates in hybrid superconducting architectures.

The present approach possesses several attractive features for scalable superconducting quantum computing. First, only a single flux-control parameter is required throughout the gate operation, reducing hardware complexity compared with multidimensional control protocols. Second, the Fourier-cosine pulse parameterization provides a low-dimensional yet sufficiently expressive control landscape, enabling efficient numerical optimization while remaining compatible with experimentally generated flux waveforms. Third, the optimization framework is independent of the specific circuit parameters and can therefore be readily extended to other tunable-coupler architectures, including transmon-transmon, fluxonium-fluxonium, and other hybrid superconducting systems. Finally, although the present work focuses on coherent Hamiltonian dynamics, the proposed optimization framework can be naturally generalized to incorporate realistic decoherence mechanisms, parameter uncertainties, control distortions, and hardware constraints. The combination of physically motivated cost functions with low-dimensional pulse parameterization may therefore provide an efficient route toward robust optimal control of multiqubit superconducting quantum processors.

Overall, the proposed hybrid fluxonium-transmon architecture, together with the Fourier-cosine optimal-control protocol, provides a practical and scalable approach for implementing high-fidelity entangling gates in superconducting quantum circuits. We expect that the present framework can be readily integrated with experimentally accessible tunable-coupler platforms and serve as a useful building block for fault-tolerant superconducting quantum information processing.

\emph{Acknowledgements}-P. X. express gratitude to Yu-Mei Song for her steadfast support throughout my academic journey. This work was financially supported by the National Natural Science Foundation of China (Grant Nos. 12105146, 92565111), the Program of State Key Laboratory of Quantum Optics Technologies and Devices (No: KF202505). Shengjun Wu is supported by the National Natural Science Foundation of China (Grant No. 12475020), the National Key Research and Development Program of China (Grant No. 2023YFC2205802), and the Innovation Program for Quantum Science and Technology (Grant No. 2021ZD0301701). Xiaohong Yan acknowledges funding from the National Natural Science Foundation of China (Grant No. 12574266) and National Key Research and Development Program of China (Grant No. 2022YFA1405200).

\appendix

\section{Perturbative Analysis of the Residual ZZ Interaction}
\label{app:analyticalZZ}

To provide an analytical description of the residual $\rm{ZZ}$ interaction in the proposed hybrid TCF architecture, we develop a perturbative treatment of the static ZZ interaction induced by the direct and coupler-mediated couplings. The perturbative analysis complements the numerical diagonalization presented in the main text and provides a microscopic interpretation of the flux dependence and suppression of the residual $\rm{ZZ}$ interaction.

Throughout this Appendix, the basis states are ordered as $\ket{\rm{TFC}}$, corresponding to the transmon, fluxonium, and coupler degrees of freedom, respectively. The coupler is assumed to remain in its ground state, so that the computational subspace is spanned by ${|000\rangle,|010\rangle,|100\rangle,|110\rangle}$.
The static $\rm{ZZ}$ interaction is defined from the dressed computational-state energies as
\begin{equation}
\zeta = E_{110}-E_{100}-E_{010}+E_{000},
\label{eq:ZZ_definition_app}
\end{equation}
where $E_{n_{\rm T}n_{\rm F}n_{\rm C}}$ denotes the dressed energy continuously connected to the corresponding bare state. Following the perturbative treatment of static $\rm{ZZ}$ coupling developed in Refs.~\cite{RKrishnanIntJQuantumChem14911978, PengZhaoPRAp0240372021, LingJiangPRA1110326092025, PengXuPRApp230340362025}, we decompose the conditional interaction into contributions of different perturbative orders,
\begin{equation}
\zeta =\zeta^{(2)} + \zeta^{(3)} + \zeta^{(4)} +\cdots,
\label{eq:ZZ_total}
\end{equation}
where
\begin{equation}
\zeta^{(n)} = E_{110}^{(n)} - E_{100}^{(n)}-E_{010}^{(n)}+E_{000}^{(n)}
\label{eq:zeta_n}
\end{equation}
is the $n$th-order contribution to the residual $\rm{ZZ}$ interaction. Separate the Hamiltonian into an unperturbed part and a perturbation,
\begin{equation}
H=H_0+H_{int}.
\end{equation}
where $H_0$ contains the individual Hamiltonians of the transmon, fluxonium, and tunable coupler, whereas $H_{int}$ describes their mutual coupling, see Eq.(\ref{Hint}).

The perturbative corrections to the energy of an unperturbed state $\ket{s}$ are given by
\begin{equation}
E_s^{(2)} = \sum_{j\neq s} \frac{|V_{sj}|^2}{E_s^{(0)}-E_j^{(0)}},
\label{eq:E2_app}
\end{equation}
\begin{equation}
E_s^{(3)} =\sum_{j,k\neq s} \frac{V_{sj}V_{jk}V_{ks}}{\left(E_s^{(0)}-E_j^{(0)}\right)\left(E_s^{(0)}-E_k^{(0)}\right)},
\label{eq:E3_app}
\end{equation}
and
\begin{equation}
\begin{aligned}
E_s^{(4)} ={} \sum_{j,k,l\neq s} \frac{V_{sj}V_{jk}V_{kl}V_{ls}}{E_{sj}E_{sk}E_{sl}} -\sum_{j,k\neq s}\frac{|V_{sj}|^2|V_{sk}|^2}{E_{sj}^{\,2}E_{sk}},
\end{aligned}
\label{eq:E4_app}
\end{equation}
where
\begin{equation}
V_{sj}=\langle s|V|j\rangle,
\qquad
E_{sj}=E_s^{(0)}-E_j^{(0)}.
\end{equation}

Because the fluxonium is strongly anharmonic, its transition matrix elements cannot in general be expressed by the harmonic-oscillator relation $\sqrt{n+1}$. We therefore retain the relevant matrix elements explicitly. We define
\begin{equation}
\rm{T}_{01} =\left|\langle 0_{\rm{T}}|\hat{\mathit{a}}|1_{\rm{T}}\rangle\right|,
\rm{F}_{01} =\left|\langle 0_{\rm{F}}|\hat{\mathit{b}}|1_{\rm{F}}\rangle\right|,
\rm{F}_{12} =\left|\langle 1_{\rm{F}}|\hat{\mathit{b}}|2_F\rangle\right|,
\label{eq:app_matrix_F}
\end{equation}
and
\begin{equation}
\rm{C}_{01}=\left|\langle 0_{\rm{C}}|\hat{\mathit{c}}|1_{\rm{C}}\rangle\right|.
\label{eq:app_matrix_C}
\end{equation}
For the weakly anharmonic transmon and tunable coupler, the higher transition matrix elements are approximated by the corresponding harmonic-oscillator scaling, such that
\begin{equation}
\left|\langle 1_{\rm{T}}|\hat{\mathit{a}}^\dagger|0_{\rm{T}}\rangle\right|=\rm{T}_{01},
\qquad
\left|\langle 2_{\rm{T}}|\hat{\mathit{a}}^\dagger|1_{\rm{T}}\rangle\right|=\sqrt{2}\,\rm{T}_{01},
\end{equation}
and similarly for the coupler.

To make the flux dependence explicit, we introduce the detunings
\begin{equation}
\Delta_{\rm{TF}} = \omega_{\rm T}-\omega_{\rm F},
\label{eq:DeltaTF_app}
\end{equation}
\begin{equation}
\Delta_{\rm T}(\Phi_{ext}) =\omega_{\rm T} - \omega_{\rm C}(\Phi_{ext}),
\label{eq:DeltaT_app}
\end{equation}
and
\begin{equation}
\Delta_{\rm F}(\Phi_{ext}) = \omega_{\rm F} - \omega_{\rm C}(\Phi_{ext}),
\label{eq:DeltaF_app}
\end{equation}
where $\omega_{\rm C}(\Phi_{ext})$ is the flux-dependent transition frequency of the tunable coupler.

\begin{figure}
\begin{center}
\includegraphics[width = 7.50cm, height = 5.50cm]{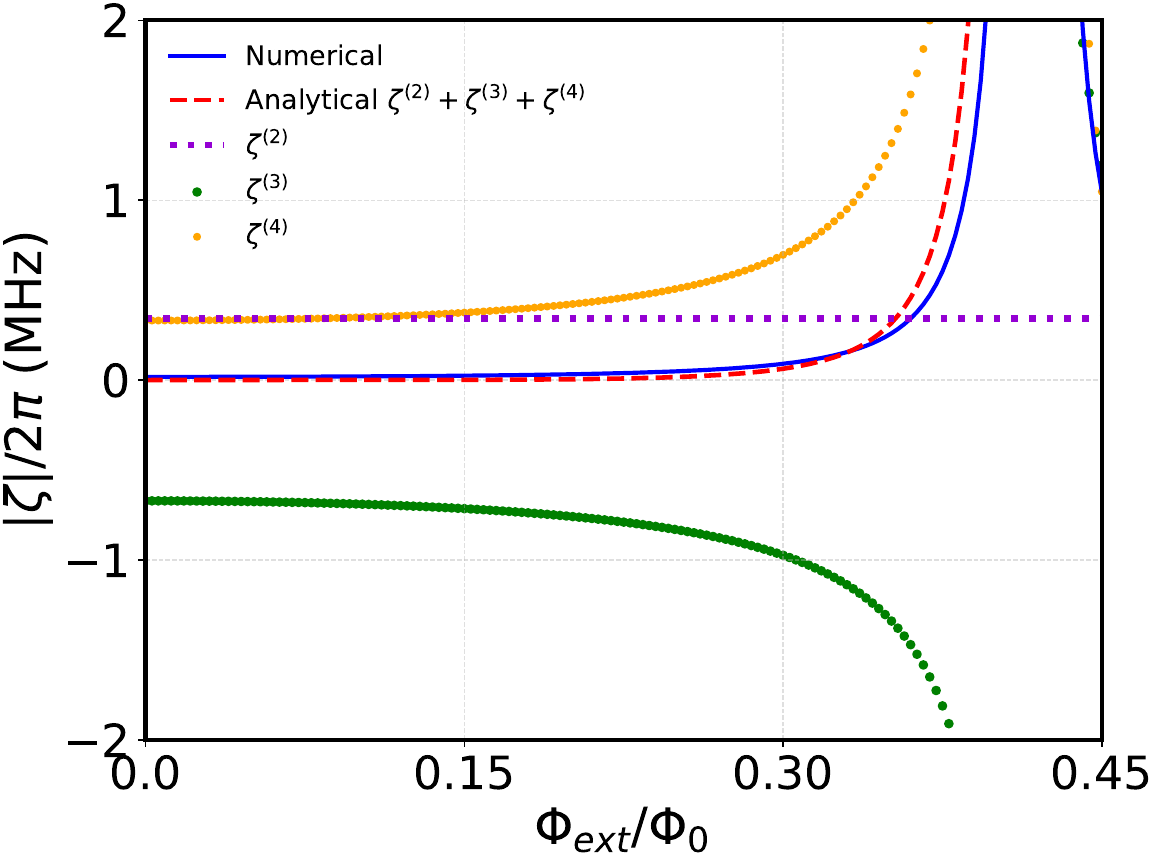}
\end{center}
\caption{Residual $\rm{ZZ}$ interaction obtained from perturbative and exact numerical calculations as a function of the coupler flux bias.}
\label{Fig7ZZAnaNum}
\end{figure}


After combining the second-order corrections to the four computational-state energies, the leading direct-coupling contribution becomes
\begin{equation}
\zeta^{(2)} = \frac{\left(J_{\rm TF}{\rm{T}}_{01}{\rm{F}}_{12}\right)^2}{\Delta_{\rm TF}-\alpha_{\rm{F}}}
-\frac{\left(J_{\rm TF}\sqrt{2}\,{\rm{T}}_{01}{\rm{F}}_{01}\right)^2}{\Delta_{\rm TF}+\alpha_{\rm{T}}}.
\label{eq:app_zeta2}
\end{equation}
This term is independent of the coupler flux to leading order and therefore represents the residual static interaction associated with the direct $\rm T$-$\rm F$ coupling. In particular, the two terms originate from virtual transitions through the higher transmon and fluxonium levels, respectively. Owing to the strong anharmonicity of the fluxonium, retaining the distinct matrix elements ${\rm{F}}_{01}$ and ${\rm{F}}_{12}$ is important for quantitatively describing the residual ZZ interaction.


Using the corresponding transition matrix elements, the third-order contribution to the ZZ interaction can be written as
\begin{align}
\zeta^{(3)} ={}&\frac{2\left(J_{\rm TC}{\rm{C}}_{01}{\rm{T}}_{01}\right)\left(J_{\rm FC}{\rm{C}}_{01}{\rm{F}}_{12}\right)\left(J_{\rm TF}{\rm{T}}_{01}{\rm{F}}_{12}\right)}{\left(\Delta_{\rm TF}-\alpha_{\rm{F}}\right)\Delta_{\rm{T}}}\nonumber\\
&-\frac{2\left(J_{\rm TC}{\rm{C}}_{01}\sqrt{2}\,{\rm{T}}_{01}\right)\left(J_{\rm FC}{\rm{C}}_{01}{\rm{F}}_{01}\right)\left(J_{\rm TF}\sqrt{2}\,{\rm{T}}_{01}{\rm{F}}_{01}\right)}{\left(\Delta_{\rm TF}+\alpha_{\rm{T}}\right)\Delta_{\rm{F}}}\nonumber\\
&+\frac{4\left(J_{\rm FC}{\rm{C}}_{01}{\rm{F}}_{01}\right)\left(J_{\rm TC}{\rm{C}}_{01}{\rm{T}}_{01}\right)\left(J_{\rm TF}{\rm{T}}_{01}{\rm{F}}_{01}\right)}{\Delta_{\rm{T}}\Delta_{\rm{F}}}.
\label{eq:app_zeta3}
\end{align}
The three terms correspond to distinct virtual-transition pathways connecting the transmon, fluxonium, and coupler. Importantly, because the coupler transition frequency $\omega_{\rm{C}}$ is flux tunable, both $\Delta_{\rm{T}}$ and $\Delta_{\rm{F}}$ depend on the external flux bias. Consequently, $\zeta^{(3)}$ provides a flux-dependent contribution to the effective ZZ interaction.


Keeping the leading coupler-mediated fourth-order pathways gives
\begin{align}
\zeta^{(4)}& =\frac{\left(J_{\rm TC}{\rm{C}}_{01}{\rm{T}}_{01}\right)^2\left(J_{\rm FC}{\rm{C}}_{01}{\rm{F}}_{12}\right)^2}{\Delta_{\rm{T}}^2\left(\Delta_{\rm TF}-\alpha_{\rm{F}}\right)}\nonumber\\
&-\frac{\left(J_{\rm TC}{\rm{C}}_{01}\sqrt{2}\,{\rm{T}}_{01}\right)^2\left(J_{\rm FC}{\rm{C}}_{01}{\rm{F}}_{01}\right)^2}{\Delta_{\rm{F}}^2\left(\Delta_{\rm TF}+\alpha_{\rm{T}}\right)}\nonumber\\
&+\frac{\left(J_{\rm TC}{\rm{C}}_{01}{\rm{T}}_{01}\right)^2\left(J_{\rm FC}\sqrt{2}\,{\rm{C}}_{01}{\rm{F}}_{01}\right)^2}{\left(\Delta_{\rm{T}}+\Delta_{\rm{F}}-\alpha_{\rm{C}}\right)\Delta_T^2}\nonumber\\
&+\frac{\left(J_{\rm TC}{\rm{C}}_{01}\sqrt{2}\,{\rm{T}}_{01}\right)^2\left(J_{\rm FC}{\rm{C}}_{01}{\rm{F}}_{01}\right)^2}{\left(\Delta_{\rm{T}}+\Delta_{\rm{F}}-\alpha_{\rm{C}}\right)\Delta_{\rm{F}}^2}\nonumber\\
&+\frac{2{\sqrt{2}\left(J_{\rm TC}{\rm{C}}_{01}{\rm{T}}_{01}\right)}^2{\sqrt{2}\left(J_{\rm FC}{\rm{C}}_{01}{\rm{F}}_{01}\right)}^2}{\left(\Delta_{\rm{T}}+\Delta_{\rm{F}}-\alpha_{\rm{C}}\right)\Delta_{\rm{T}}\Delta_{\rm{F}}}.
\label{eq:app_zeta4}
\end{align}

The first two terms contain the direct transmon-fluxonium detuning in their denominators and describe fourth-order corrections to the virtual excitation pathways involving the second excited states of the fluxonium and transmon, respectively. The last three terms arise from virtual excitation of the coupler and contain the energy denominator $\Delta_{\rm{T}}+\Delta_{\rm{F}}-\alpha_{\rm{C}}$,
which reflects the relevant higher excited state of the coupler.


Combining the above contributions, the residual ZZ interaction is approximated by
\begin{equation}
\rm ZZ_{\rm pert} = \zeta^{(2)} + \zeta^{(3)} + \zeta^{(4)}.
\label{eq:app_zeta_total}
\end{equation}
The perturbative expression highlights the distinct physical mechanisms contributing to the residual ZZ interaction. The corresponding quantity plotted in the Fig~\ref{Fig7ZZAnaNum}. The second-order term originates from the direct $\rm T$-$\rm F$ coupling and is therefore approximately flux independent. In contrast, the third- and fourth-order terms involve the tunable coupler and acquire a pronounced dependence on the external flux through $\omega_C(\Phi_{ext})$, $\Delta_{\rm{T}}$, and $\Delta_{\rm{F}}$. As the coupler is tuned toward the interaction region, the corresponding virtual-transition pathways become increasingly important and modify the effective $\rm T$-$\rm F$ interaction through level repulsion and interference between different virtual processes.

The perturbative results agree well with the exact numerical calculations in the large-detuning regime, but deviate increasingly near resonance, where strong level hybridization invalidates the weak-coupling assumption underlying the perturbative expansion. The decomposition into different orders further shows that $\zeta^{(2)}$ is nearly flux independent, providing a background contribution, whereas the third- and fourth-order terms exhibit pronounced flux dependence due to coupler-mediated virtual processes. Importantly, these higher-order contributions undergo destructive interference over a suitable flux range, leading to a substantial suppression of the total residual $\rm{ZZ}$ interaction. Thus, flux tuning enables the compensation of the approximately flux-independent background by the flux-dependent higher-order contributions.

\begin{figure}
\begin{center}
\includegraphics[width = 7.50cm, height = 5.50cm]{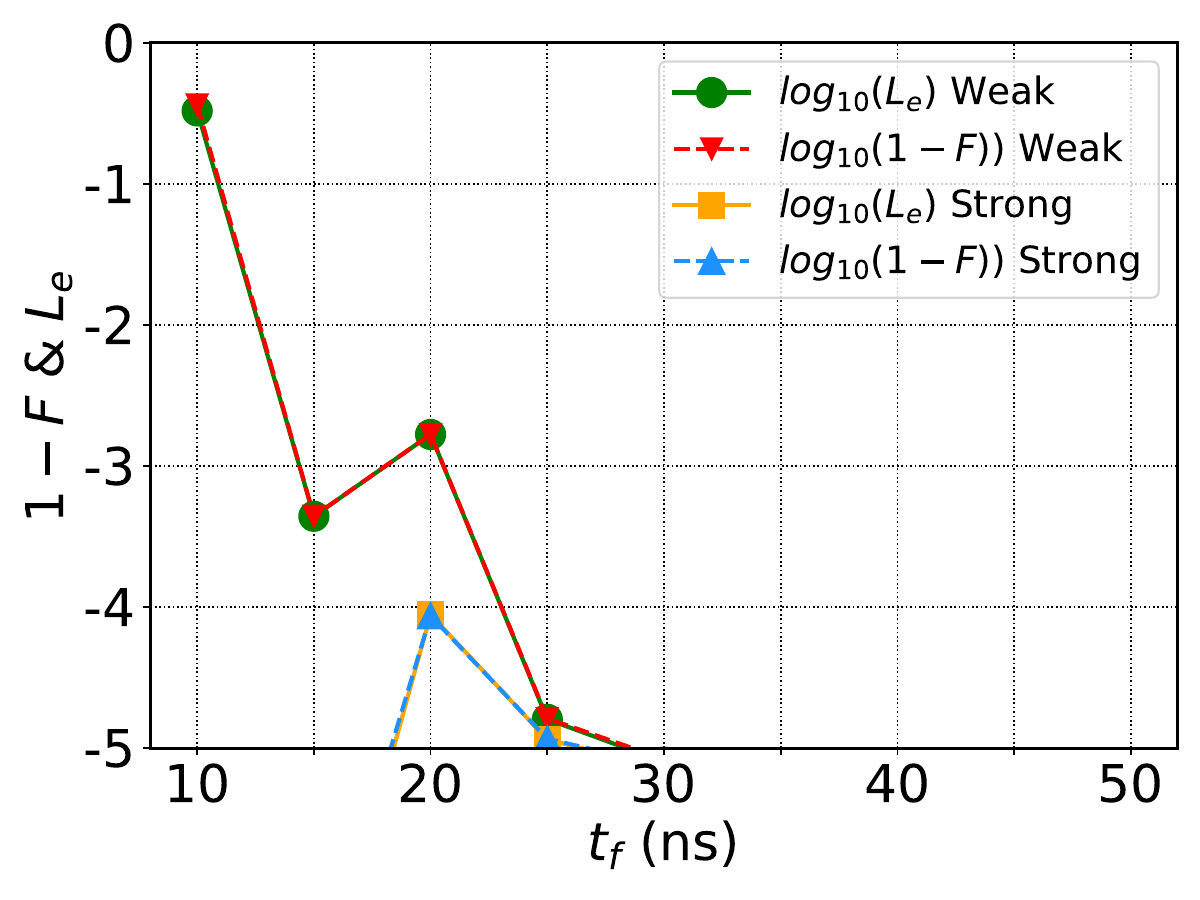}
\end{center}
\caption{Comparison of the optimized CZ-gate performance for weak- and strong-coupling configurations as a function of the gate duration. The results illustrate the trade-off between gate speed, leakage, and coherent control errors associated with different coupling strengths.}
\label{Fig8}
\end{figure}

\section{Gate-Time Dependence under Different Coupling Strengths}
\label{app:strongcoupling}

The coupling strength between superconducting qubits is an important parameter in determining the characteristic timescale of an entangling gate. In general, stronger interactions can accelerate population transfer and conditional-phase accumulation, thereby enabling faster gate operations. However, excessively strong coupling may also enhance unwanted transitions, leakage, and residual interactions. To clarify this trade-off, we compare the gate performance under weak- and strong-coupling configurations over a broad range of gate durations.

For the strong-coupling configuration, we increase the interaction strengths to $J_{\rm{TC}}/2\pi=100$~MHz, $J_{\rm{FC}}/2\pi=500$~MHz, and $J_{\rm{TF}}/2\pi=50$~MHz, while keeping the other system parameters unchanged. The same flux-pulse optimization procedure is then applied over the same range of gate durations considered for the weak-coupling configuration. The numerical results show that the enhanced interactions substantially reduce the minimum gate time required for high-fidelity CZ-gate operation. In particular, a gate fidelity exceeding $99.999\%$ can be achieved within approximately $10$~ns in the strong-coupling regime. The corresponding gate infidelity and leakage errors for the two coupling configurations are shown in Fig.~\ref{Fig8}, where the different operating regimes associated with weak and strong coupling can be clearly identified.

A direct comparison between the two coupling regimes reveals a clear dependence of the optimal coupling strength on the gate-time regime. For short gate durations, the stronger interactions are advantageous because they accelerate the desired population transfer and conditional-phase accumulation, allowing the CZ operation to be completed before decoherence becomes significant. As the gate duration is increased, however, the stronger coupling also enhances the participation of noncomputational states and residual coupler-mediated interactions. These effects can increase leakage and coherent control errors and consequently limit the achievable gate fidelity.

In contrast, the weak-coupling configuration provides a more favorable operating regime for relatively long gate durations. The reduced interaction strength suppresses unwanted transitions and residual interactions, while the longer control time provides greater freedom for shaping the flux pulse and suppressing nonadiabatic leakage. Consequently, the weak-coupling configuration can achieve improved intrinsic gate fidelity as the gate duration is increased.

These results indicate that there is no universally optimal coupling strength independent of the gate-time regime. Strong coupling is particularly advantageous for ultrafast CZ gates, whereas weaker coupling provides greater flexibility for longer-duration and more adiabatic operations. When decoherence is taken into account, the balance shifts toward shorter gate times because prolonged gate evolution increases exposure to energy relaxation and dephasing. Therefore, an appropriately engineered and controllable coupling strength provides a practical compromise among gate speed, leakage suppression, coherent-error control, and decoherence, enabling high-fidelity CZ-gate operation across a broad range of gate durations.

\end{document}